\documentclass[aps,pre,twocolumn,amsmath,amssymb,superscriptaddress,longbibliography]{revtex4-2}

\usepackage{mwe}
\usepackage{graphicx}
\usepackage{dcolumn}
\usepackage{bm}
\usepackage{hyperref}

\begin{document}

\preprint{APS/123-QED}

\title{Length-scale selection in adaptive transport networks}

\author{Sidney Holden}
\affiliation{Center for Computational Biology, Flatiron Institute, New York, NY 10010, USA}

\author{Mia C. Morrell}
\affiliation{Department of Physics and Center for Soft Matter Research,
New York University, New York, NY, 10003, USA}
\affiliation{Center for Computational Biology, Flatiron Institute, New York, NY 10010, USA}

\author{Geoffrey Vasil}
\affiliation{School of Mathematics and the Maxwell Institute for Mathematical Sciences, University of Edinburgh, Edinburgh,
UK, James Clerk Maxwell Building, Edinburgh, EH9 3FD, UK}

\author{Eleni Katifori}
\affiliation{Center for Computational Biology, Flatiron Institute, New York, NY 10010, USA}
\affiliation{Department of Physics and Astronomy, University of Pennsylvania,
Philadelphia PA, 19104, USA}

\begin{abstract}
Adaptive transport networks in biological and physical systems exhibit
hierarchical organization, characteristic channel spacing, and robust
scaling relations. 
Existing adaptive network models, formulated on a lattice, successfully reproduce many observed topologies and conduit scaling laws; however, the mechanism that selects network density and spatial spacing remains unclear.
We address this in a continuum formulation where conductivity
evolves as a tensor field coupled to pressure-driven flow. Linearizing
about a homogeneous conducting state, we identify a finite-wavelength
instability with preferred wavelength scaling $\lambda \sim \mathcal{R}^{-1/4}$ in the control parameter $\mathcal{R}$. Simulations of the full equations confirm the analytical predictions and demonstrate the formation of anisotropic conducting structures above threshold. These results establish a scale-selection principle for adaptive transport network formation which arises from a pattern-forming instability rather than solely from relaxation within a nonconvex energy landscape. 
The instability mechanism places adaptive transport systems within a broader class of nonequilibrium pattern-forming media in which constitutive transport feedback generates spatial organization. Beyond reproducing hierarchical scaling laws, the theory additionally predicts the intrinsic density of transport networks and the spatial scale of resource delivery.
\end{abstract}

\maketitle 

Transport networks typically distribute resources through hierarchical, self-similar architectures across a range of scales \cite{West1997, Banavar1999}.
Despite enormous diversity in substrate and function, these networks share a characteristic spacing between channels at each hierarchical level and power-law relations between parent and daughter conduits, of which Murray’s law in vasculature and other systems is the canonical example \cite{Murray1926, McCulloh2003}.
A widely adopted approach for modeling their formation and adaptive strategies minimizes a discrete network energy of an initially uniform lattice subject to loading--a particular example of physical learning \cite{Stern2023}.
This has reproduced many topological features and scaling laws found in real networks, from leaf venation \cite{Sack2013, Katifori2010, Matos2025, Alonso2025} and fungal networks \cite{Tero2010, Schick2024, OyarteGalvez2025} to animal vasculature \cite{Secomb2017, Kirkegaard2020, Qi2024} and river systems \cite{Sinclair1996, Rinaldo2014, Konkol2022}.
A classic model--first employed to describe physarum \emph{solving} \cite{Tero2006}--tunes edge weights $K_e$ (conductivities of edge $e$) according to the flow $Q_e$ they carry, following a ``use-it-or-lose-it'' local rule. 
Hu and Cai \cite{HuCai2013} derived this rule from a cost function-minimization principle that includes the total pumping power needed, $\sum_e Q_e \Delta P_e$, as well as the associated metabolic cost, $\sum_e K_e^{\gamma}$, with the metabolic exponent $\gamma$ distinguishing morphologies.

Despite the model successes in reproducing many aspects of transport network morphology, what morphologically sets the observed characteristic channel spacing remains poorly understood. 
Vein densities in leaves \cite{Noblin2008, Sack2013, He2024} and capillary densities in vascular beds \cite{Secomb2017} set the spatial frequency of supply within a given tissue, but vary by orders of magnitude across species and physiological conditions. From a functional perspective such variations in vascular density can be explained by the different functional demands of each tissue, including the different oxygenation needs of muscle and brain \cite{Goldman2008}, or the vein to stoma water evaporation pathways in leaves \cite{Noblin2008}. However, the need remains for a simple morphodynamic model that can select a characteristic lengthscale for the vein density. 

A continuum approach is the natural setting to understand the spatial scale-selection, with the powerful tools of linear instability theory and asymptotic analysis.

The idea is to replace pressure at nodes with a continuous scalar field $p$, and discrete edges with a conductivity tensor field, $K$, allowing networks to emerge without prescribing an underlying graph \cite{HuCai2019, Haskovec2015, Burger2019}.
Numerical and analytical studies of the continuum vessel adaptation, morphogenetic equations have revealed a rich range of behaviors, including filament formation, tensorial effects, mesoscopic limits, and phase-field extensions \cite{Albi2016,Albi2017,Haskovec2019ODEPDE,Haskovec2019b,Astuto2022,Astuto2023,Xia2023,Astuto2024,Astuto2024b,Li2024,Haskovec2025}.

\begin{figure*}
    \centering
    \includegraphics[width=\textwidth]{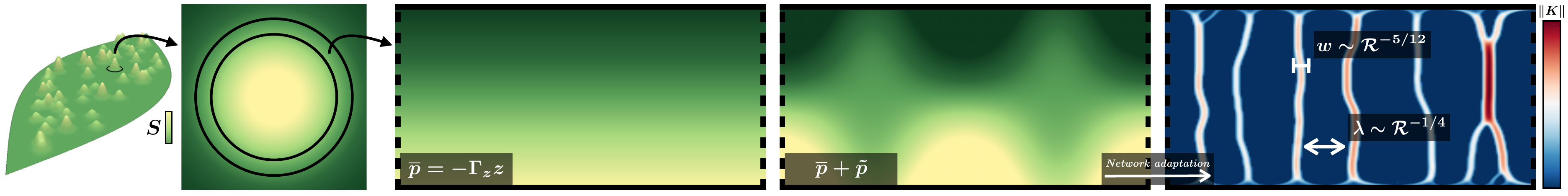}
\caption{Schematic of the local stability framework. The full network problem (left) features a heterogeneous source $S$. In a small bulk patch (annulus) where $S \approx 0$, the leaf-scale flow appears as a uniform background pressure gradient $\nabla \bar{p} = -\Gamma_z \hat{\bm{e}}_z$, imposed via periodic-in-$x$, prescribed-flux-in-$z$ boundary conditions, with homogeneous base-state conductivity $\bar{K} = \bar{K}_{zz}\,\hat{\bm{e}}_z \otimes \hat{\bm{e}}_z$. Linearization yields a band of unstable modes [Fig.~\ref{fig:linear_theory}(a)]; saturation and coarsening reorganize $\|K\|$ into a network of vertically oriented channels (right) of width $w \sim \mathcal{R}^{-5/12}$ and spacing $\lambda \sim \mathcal{R}^{-1/4}$, set by the competition in Eq.~\eqref{eq:adapt_cont} between transport-driven growth ($\nabla p \otimes \nabla p$), metabolic decay ($f(\|K\|)\,K$), and diffusive smoothing ($\mathcal{R}^{-1}\nabla^2 K$), subject to current conservation [Eq.~\eqref{eq:poisson}].}
\label{fig:schematic}
\end{figure*}

Here, we show that a continuum version of the adaptive equations possesses a finite-wavelength instability of a spatially homogeneous conducting state.
To isolate the mechanism selecting the characteristic spacing and orientation of these structures, in this work we consider a uniform background maintained by a constant through-flow and derive the linearized dynamics of coupled perturbations in pressure and conductivity.
We show how the resulting dispersion relation yields both a critical control parameter and a preferred wavenumber for onset. 
Above the threshold, the nonlinear network spacing is selected by the fastest-growing linear instability.
The predicted onset agrees with nonlinear simulations of the full system, demonstrating that network formation arises from a continuum instability rather than solely from the nonlinear relaxation of a nonconvex energy landscape. The mechanism shares structural features with classical finite-wavelength instabilities, including Turing-like systems \cite{Turing1952} and hydrodynamic instabilities such as Rayleigh–Bénard convection \cite{chandra1961}.
This places adaptive transport networks within the broader class of nonequilibrium systems in which transport laws and constitutive feedback generate spatial structure \cite{Cross1993}.
While adaptive transport models generate realistic networks, the instability identified here additionally specifies their characteristic spacing, complementing the parent–daughter scaling laws with a prediction for network density.

\textit{Continuum model and instability}—The discrete model produces conductance solutions with a range of structures–e.g. tree-like vs. mesh-like, hierarchical vs. uniform.
The selected structure depends on the type, distribution, and time-dependence of sources, the domain geometry and boundary conditions, and the initial conditions.

Here we isolate the network-scale selection mechanism by investigating the stability of a constant flow far from sources where $S=0$ (see Fig.~\ref{fig:schematic}), and consider a 2d periodic in $x$ rectangular strip with a background pressure gradient imposed through boundary conditions, as in figure~\ref{fig:schematic}.
Specifically, we set the background solution of the governing adaptation equations, derived from gradient descent of an energy cost function (see appendix C), as
\begin{align}
\nabla \bar{p} &= -\bar{\Gamma}_{z} \, e_{z},
&
\bar{K} &= \bar{K}_{zz}\, e_{z} \otimes e_{z}.
\end{align}
As $\bar{K}$ and $\bar{p}$ need to satisfy the governing equations, the constants $\bar{\Gamma}_{z}$ and $\bar{K}_{zz}$ satisfy $\bar{\Gamma}_{z}^{2} = \nu\,\bar{K}_{zz}^{\gamma - 1}$, where $\nu$ controls the rate of nonlinear feedback from conductivity growth. The $\gamma$ exponent governs much of the overall dynamics, with the biologically relevant regime corresponding to $1/2 \le \gamma \le 1$.
We nondimensionalize all gradients of the governing equations with the largest vertical wavenumber in the $e_{z}$ direction, $L_{z}^{-1} = \pi/H$, where $H$ is the domain thickness. We scale pressure by $L_{z}\bar{\Gamma}_{z}$, conductivity by $\bar{K}_{zz}$, and time by $\bar{K}_{zz}/\bar{\Gamma}_{z}^{2}$. 
The governing equations become
\begin{align}
& -\nabla \cdot \left( (K + \kappa I) \cdot \nabla p \right) \,=\, 0,
\label{eq:poisson}
\\
& \partial_{t}K 
\, = \, 
\nabla p \otimes \nabla p 
- f(\|K\|)\,K
+
\frac{1}{\mathcal{R}}\nabla^{2}K, 
\label{eq:adapt_cont}
\end{align}
where $\|K\|^{2} = \text{tr}(K^{\top}\!\!\cdot\! K)$,  $f(q) = (\tfrac{q^{2}+\varepsilon}{1+\varepsilon})^{\frac{\gamma-2}{2}}$, with $\varepsilon$ as a nonlinear regularization parameter. The parameter $\kappa$ is the homogeneous, isotropic ambient pressure diffusivity. 
The overall stability is governed by the nondimensional control parameter
\begin{align}
\mathcal R = \frac{\bar{\Gamma}_z^{2}\, L_z^{2}}{\eta \, \bar{K}_{zz}},
\end{align}
which measures the strength of conductivity growth driven by transport relative to diffusive network smoothing, $\eta$. 
We seek perturbations about the background in the form
\begin{align}
p &= -z + \tilde{p}, \\
K &= \tilde{K}_{xx}e_{xx}+ \tilde{K}_{xz}e_{xz} +  (1+\tilde{K}_{zz})\,e_{zz},
\end{align}
where $e_{ij} = \tfrac{1}{2}(e_{i} \otimes e_{j} + e_{j} \otimes e_{i})$.
The $\tilde{K}_{xx}$ component decouples linearly and decays, but becomes active in a nonlinear regime. 
At $z=0,\pi$, we set $\partial_{z} \tilde{p} = \tilde{K}_{zz} = \partial_z \tilde{K}_{xz} = 0$, with periodic boundary conditions in $x$.
The system structurally resembles hydrodynamics, with $\tilde{K}_{zz}$ and $\tilde{K}_{xz}$ coupling to pressure gradients, analogously to the fluid velocity components. 
We assume normal modes of the form $\tilde{p} = P \cos(n z) \cos(kx) e^{\sigma t}, \tilde{K}_{xz} = U_{x} \cos(nz) \sin(kx) e^{\sigma t}$ and $\tilde{K}_{zz} = U_{z} \sin(nz) \cos(kx) e^{\sigma t}$, with horizontal and vertical wave numbers, $k , n$, and growth rate, $\sigma$. 
The dispersion relation follows from the determinant of the linearized \eqref{eq:poisson}-\eqref{eq:adapt_cont},  
\begin{align}
\label{eq:dispersion}
\left| 
\begin{array}{ccc} 
\mathcal{L} & k & n \\ 
-k & \mathcal{D} & 0 \\ -2n & 0 & \mathcal{D} -(\tilde{\gamma}_{\varepsilon} +1)
\end{array} 
\right| = 0, 
\end{align} 
where $\mathcal{L} = \kappa (k^{2} +n^{2}) + n^{2}$, $\mathcal{D} = \sigma + \tfrac{k^{2} + n^{2}}{\mathcal{R}} + 1 $, and $\tilde{\gamma}_{\varepsilon} = \frac{1-\gamma-\varepsilon}{1+\varepsilon}$ (see appendix \ref{app:derivations}).
The vertical wave number is a positive integer, $n = 1,2,\ldots$.
The eigenmode structure arising from the full Fourier expansion in $x$ shows that the only unstable modes are for real $\sigma$. The mode polarization is 
\begin{align}
    (P,U_{x},U_{z}) &\propto (-n \mathcal{D},\, -kn,\,  k^{2} + \mathcal{L}\,\mathcal{D}).
\end{align}
The linear mode amplitude is defined only up to an arbitrary overall normalization. 

\begin{figure}
    \centering
    \includegraphics[width=\columnwidth]{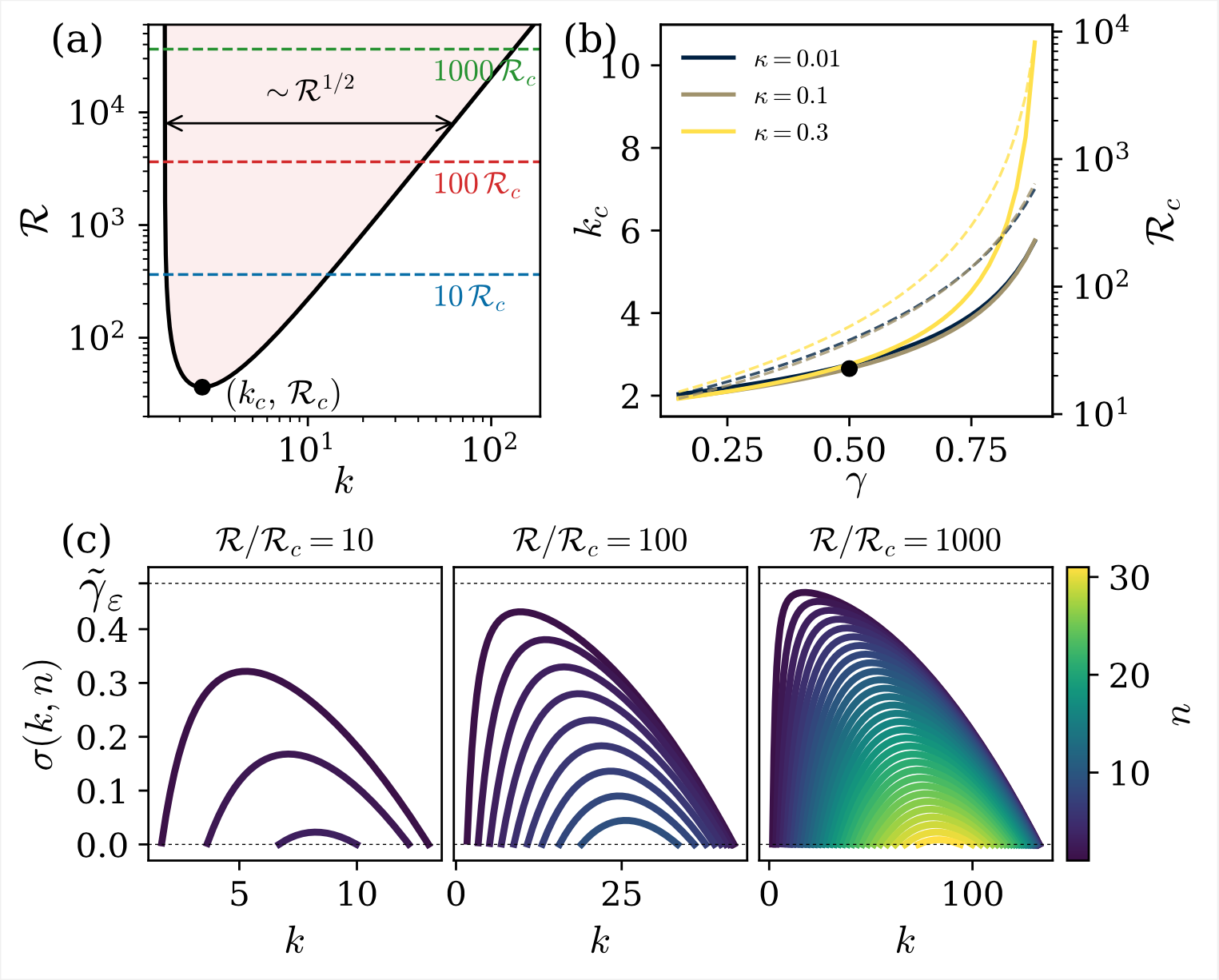}
\caption{Linear theory for the continuum adaptive transport model. 
(a) Neutral stability curve $\mathcal{R}(k)$ for vertical mode $n=1$ at $\gamma = 0.5$, $\kappa = 0.1$; shading marks the unstable region, with critical point $(k_c, \mathcal{R}_c) \approx (2.655, 36.39)$. Horizontal dashed lines: supercriticalities $\mathcal{R}/\mathcal{R}_c \in \{10, 100, 1000\}$. 
(b) Critical wavenumber $k_c$ (solid, left axis) and critical control parameter $\mathcal{R}_c$ (dashed, right axis) versus $\gamma$, for three values of $\kappa$; the operating point $(\gamma, \kappa) = (0.5, 0.1)$ is marked. 
(c) Linear growth rates $\sigma(k,n)$ as a function of horizontal wavenumber $k$ for vertical mode numbers $n = 1, 2, \ldots$ (color), at three supercriticalities; only unstable modes ($\sigma > 0$) are shown. The dotted line at $\tilde{\gamma}_\varepsilon = (1 - \gamma - \varepsilon)/(1 + \varepsilon)$ is the asymptotic growth-rate bound for $\mathcal{R} \gg \mathcal{R}_c$.}
\label{fig:linear_theory}
\end{figure}

\textit{Neutral stability curves}---Two natural derivations follow from the dispersion relation: setting $\sigma=0$ gives the neutral curve $\mathcal{R}(k)$, whose minimum is the critical pair $(\mathcal{R}_{\mathrm c},k_{\mathrm c})$ (figure~\ref{fig:linear_theory}(a)); fixing $\mathcal{R}$ and optimizing over $k$ gives the fastest-growing mode at that supercriticality (figure~\ref{fig:linear_theory}(c)).
For illustration, the problem admits closed-form expressions when $\kappa=0$. 
The minimum of the critical curve happens at
\begin{align}
k_{\mathrm c}^{2}
&=
\frac{2+(2-\tilde{\gamma}_{\varepsilon}) \sqrt{1+\tilde{\gamma}_{\varepsilon}}}{\tilde{\gamma}_{\varepsilon}},
\\
\mathcal{R}_{\mathrm c}
&=
\frac{4 + 2 \tilde{\gamma}_{\varepsilon} -\tilde{\gamma}_{\varepsilon}^{2} +4 \sqrt{1+\tilde{\gamma}_{\varepsilon} }}{\tilde{\gamma}_{\varepsilon}^{2}},
\end{align}
where we have set $n=1$.
For general $\kappa$ and fixed $\mathcal{R}\gg \mathcal{R}_{\mathrm c}$, the growth rate is given asymptotically by $\sigma \sim\tilde{\gamma}_{\varepsilon} - \frac{k^{2}}{\mathcal{R}} - \frac{2(1+\tilde{\gamma}_{\varepsilon,\kappa})}{k^{2}}$ where $\tilde{\gamma}_{\varepsilon,\kappa} = \frac{\tilde{\gamma}_{\varepsilon}+(1+\tilde{\gamma}_{\varepsilon})\kappa}{1-(1+\tilde{\gamma}_{\varepsilon})\kappa}.$
The growth rate is bounded above by $\tilde{\gamma}_{\varepsilon}$, and the fastest-growing wavenumber scales as
\begin{align}
\label{eq:fastest_growing_wavenumber_scaling}
k \sim \left[2(1+\tilde{\gamma}_{\varepsilon,\kappa})\,\mathcal{R}\right]^{1/4}.
\end{align}
The instability therefore selects a characteristic network spacing $\lambda \sim \mathcal{R}^{1/4}$ already at linear onset. Below, we show that this wavelength persists into the saturated nonlinear regime.

In figure~\ref{fig:linear_theory}(b), we compute the dispersion relation and illustrate the dependence of the instability on the metabolic exponent $\gamma$ and the ambient diffusivity $\kappa$.
For $\gamma \to 1$, both $\mathcal{R}_{\text{c}}$ and $k_{\text{c}}$ diverge, reflecting the weakening of the instability mechanism as the energy functional approaches convexity. 
Smaller $\kappa$ lowers $\mathcal{R}_{\text{c}}$, making the instability easier to trigger, since the ambient diffusivity stabilizes the homogeneous state. 
In figure~\ref{fig:linear_theory}(c), increasing $n$ shifts the neutral curve to higher $\mathcal{R}_{\text{c}}$ and higher $k_{\text{c}}$: finer vertical structure requires stronger driving but selects shorter horizontal wavelengths. 
In the biologically relevant regime $\gamma \in [0.5, 0.75]$ and $\kappa \ll 1$, the instability onset occurs at moderate $\mathcal{R}_{\text{c}}$ with a well-defined finite wavelength.

\textit{Wavelength selection across supercriticality}---The central prediction of the linear theory is that the spacing of the saturated nonlinear network is selected by the fastest-growing instability mode, with $\lambda \sim \mathcal{R}^{-1/4}$.
We test this by running simulations at $\mathcal{R}/\mathcal{R}_c = 100$--$5000$
with fixed $\gamma = 0.5$, $\kappa = 0.1$, $\varepsilon = \kappa^2$. The
critical parameters from Eq.~\eqref{eq:dispersion} are $k_c \approx 2.655$
and $\mathcal{R}_c \approx 36.39$, and we set $L_x = 8 \times 2\pi/k_c$
to include several critical wavelengths in $x$. 
The unstable band broadens in both $k$ and $n$ as $\mathcal{R}$
increases [Fig.~\ref{fig:linear_theory}(c)], allowing for
richer nonlinear dynamics at stronger supercriticality.

\begin{figure*}
    \centering
    \includegraphics[width=\textwidth]{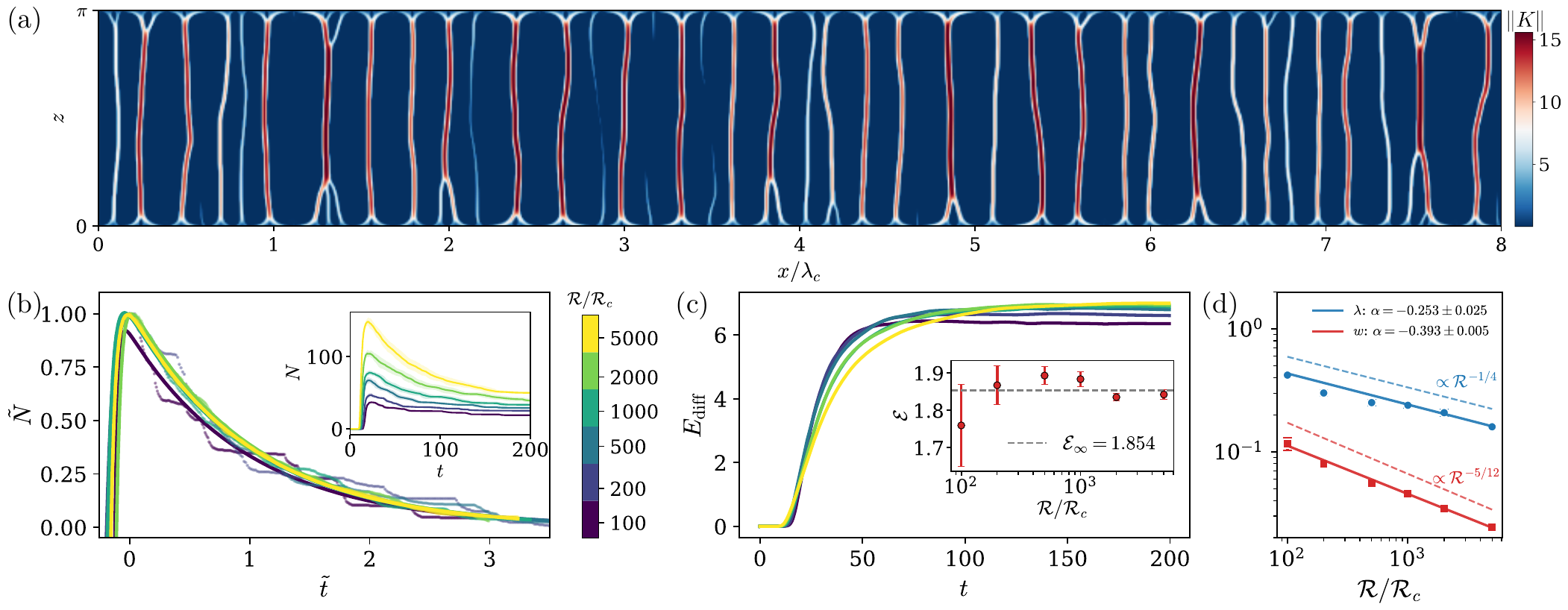}
    \caption{Filamentary saturation and asymptotic scaling.
    (a) Conductivity magnitude $\|K\|$ at $\mathcal{R}/\mathcal{R}_c = 2000$, $t = 200$; the domain spans 8 critical wavelengths
    $\lambda_c = 2\pi/k_c \approx 2.37$.
    (b) Collapse of the filament count $N(t)$ for six supercriticalities, where $\tilde{N} \equiv (N - N_{\infty}) / (N_{\text{peak}} - N_{\infty})$ and $\tilde{t} \equiv (t - t_{\text{peak}}) / \tau_{\text{coarsen}}$. Dots are data, solid lines the two-exponential fit Eq.~\eqref{eq:two_exp}. Inset: raw $N(t)$ with $\pm 1\sigma$ shading across $z$-rows.
    (c) Diffusive energy $E_\mathrm{diff}(t)$ approaching saturation. Inset: dimensionless invariant $\mathcal{E} = K_\mathrm{ch}^2/(\mathcal{R}\,w\,\lambda)$ versus $\mathcal{R}/\mathcal{R}_c$; error bars are SEM across $z$-rows. Dashed line: $\mathcal{E}_\infty \approx 1.85$, averaged over $\mathcal{R}/\mathcal{R}_c \geq 1000$.
    (d) Mean filament spacing $\lambda$ (blue) and width $w$ (red) versus $\mathcal{R}/\mathcal{R}_c$, sampled at $\tilde t = 3$ with $\pm 1\sigma$ across $z$-rows. Solid lines are power-law fits over $\mathcal{R}/\mathcal{R}_c \geq 1000$; dashed lines are $\lambda \propto \mathcal{R}^{-1/4}$ (linear theory) and $w \propto \mathcal{R}^{-5/12}$ (hypothesized).}
    \label{fig:K_and_diags}
\end{figure*}

We initialize $\tilde p$, $\tilde K$ with white noise to populate unstable modes uniformly. 
A representative saturated state at $\mathcal{R}/\mathcal{R}_c = 2000$, $t = 200$ [Fig.~\ref{fig:K_and_diags}(a)] shows the regular array of high-conductivity filaments separated by low-conductivity gaps.

To extract a quantitative measure of the dynamics we count filaments
---connected segments along which $\|K\|$ exceeds 5\% of its global
maximum---along horizontal slices through the domain, and track the
mean count $N(t)$. Every trajectory is well-described by the
two-timescale form
\begin{equation}
  N(t) = N_\infty + A\,e^{-\frac{t-t_0}{\tau_\mathrm{coarsen}}}
                  - B\,e^{-\frac{t-t_0}{\tau_\mathrm{fast}}},
  \label{eq:two_exp}
\end{equation}
with $t_0$ the onset, $\tau_\mathrm{fast}$ the rise time to peak, and
$\tau_\mathrm{coarsen}$ the post-peak relaxation. Rescaling time by
$\tilde t \equiv (t - t_\mathrm{peak})/\tau_\mathrm{coarsen}$ and
amplitude by $\tilde N \equiv (N - N_\infty)/(N_\mathrm{peak} - N_\infty)$
collapses all six trajectories onto a near-universal shape
[Fig.~\ref{fig:K_and_diags}(b)], so a single dimensionless time governs
the dynamics across two decades of supercriticality. 
The two timescales are well separated ($\tau_\mathrm{coarsen} \gg \tau_\mathrm{fast}$). 
We sample each run at $\tilde t = 3$ to compare simulations at the same dynamical stage, near the end of the accessible simulation window.

We now test whether the spacing selected at linear onset survives nonlinear saturation.
At $\tilde{t}= 3$, the mean channel spacing $\lambda = L_x/N$ and mean
width $w$ scale with supercriticality as
$\lambda \propto \mathcal{R}^{-0.251 \pm 0.006}$ and
$w \propto \mathcal{R}^{-0.392 \pm 0.017}$ in the asymptotic regime
$\mathcal{R}/\mathcal{R}_c \geq 1000$ [Fig.~\ref{fig:K_and_diags}(d)].
The spacing matches the linear prediction $\lambda \propto \mathcal{R}^{-1/4}$ to statistical precision, demonstrating that the wavelength selected at linear onset survives nonlinear saturation

Having established that linear onset predicts the nonlinear spacing, we next ask how the saturated channel amplitude scales with $\lambda \propto \mathcal{R}^{-1/4}$.
The diffusion energy
$E_\mathrm{diff} = \mathcal{R}^{-1}\!\int |\nabla K|^2\,\mathrm{d}x\,\mathrm{d}z$
rises sharply during channel formation and saturates by $t \approx 75$
at a value that is $\mathcal{R}$-independent up to logarithmic
corrections, drifting from $\approx 6.4$ at
$\mathcal{R}/\mathcal{R}_c = 100$ to $\approx 7.0$ at
$\mathcal{R}/\mathcal{R}_c = 5000$ [Fig.~\ref{fig:K_and_diags}(c)].
This near-constancy is mildly paradoxical: $E_\mathrm{diff}$ carries a
prefactor $\mathcal{R}^{-1}$ that vanishes as $\mathcal{R}\to\infty$,
and only by sharpening its gradients so that
$|\nabla K|^2 \sim \mathcal{R}$ can the system maintain a finite
diffusion energy. The diffusive term is anomalous in the sense familiar
from turbulence~\cite{Frisch1995}: it contributes a
finite share of the energy budget in the singular limit. The
dimensionless combination
\begin{equation}
  \mathcal{E} \equiv \frac{K_\mathrm{ch}^2}{\mathcal{R}\,w\,\lambda},
  \label{eq:E-invariant}
\end{equation}
constructed from the saturated center-channel conductivity
$K_\mathrm{ch}$, channel width $w$, and spacing $\lambda$, captures the
anomaly cleanly: it is $\mathcal{R}$-independent up to logarithmic
corrections, asymptoting to $\mathcal{E}_\infty \approx 1.85$
[Fig.~\ref{fig:K_and_diags}(c), inset]. With $\lambda \propto \mathcal{R}^{-1/4}$ from linear theory and
$w \propto \mathcal{R}^{-5/12}$ from the in-channel diffusion balance,
$\mathcal{E}_\infty$ fixes the channel amplitude as
$K_\mathrm{ch} \propto \mathcal{R}^{1/6}$.

\textit{Discussion}--- We have identified a mechanism of intrinsic scale selection in adaptive transport networks. The continuum transport equations possess a finite-wavelength instability whose fastest-growing mode predicts the spacing of the saturated nonlinear network. The instability provides a dynamical route to network formation that complements existing energy-minimization and relaxation-based interpretations of adaptive transport. Rather than selecting morphology solely through nonlinear relaxation within a nonconvex landscape, the system exhibits a genuine finite-wavelength pattern-forming instability with a predictable onset and characteristic wavelength.

The central claim of this work is that the linear theory of this instability predicts the saturated nonlinear dynamics. The fastest-growing wavelength $\lambda \sim \mathcal{R}^{-1/4}$ from the linear dispersion relation is the wavelength of the saturated network across more than two decades of supercriticality, surviving without coarsening to larger scales. Closure is provided by the diffusive anomaly: the regularizing diffusion contributes a finite share of the energy budget in the limit it would formally vanish, encoded in the dimensionless invariant $\mathcal{E}_\infty$. Together with the in-channel diffusion balance, this fixes the channel amplitude $K_\mathrm{ch} \propto \mathcal{R}^{1/6}$, so that the full macroscopic structure of the saturated network is a prediction of the linear instability that produces it. Whatever small-scale physics actually plays the role of $\mathcal{R}^{-1}\nabla^2 K$ in real adaptive transport networks --- finite vessel diameter, anisotropic
flux, stochastic fluctuations in supply --- a relation of this form should survive, since it reflects the macroscopic energy balance rather than the regularizer itself.

Although motivated by biological transport networks, the mechanism is more general. Similar feedbacks between transport, reinforcement, dissipation, and local smoothing appear in vascular remodeling, leaf venation, fungal transport networks, river formation, and related adaptive flow systems. In each case, local transport enhances the conductivity of existing pathways while competing processes suppress short-scale structure, naturally generating a preferred spacing between transport channels. The present analysis isolates this mechanism in its simplest continuum setting and places adaptive transport networks within the broader class of nonequilibrium finite-wavelength pattern-forming systems.

Several extensions follow naturally from this framework. Spatially varying source distributions, time-dependent forcing, growth, and curved geometries should all modify the local control parameter and therefore the selected network scale. The continuum instability identified here provides a starting point for understanding how hierarchical transport architectures emerge from spatially heterogeneous environments and evolve across scales.

\bibliographystyle{apsrev4-2}
\bibliography{references}

\newpage

\appendix

\section{Linear Theory Derivations}\label{app:derivations}

The active linearized equations are
\begin{align}
\partial_{x}\tilde{K}_{xz} + \partial_{z}\tilde{K}_{zz} 
&=
\kappa \nabla^{2}\tilde{p} + \partial_{z}^{2}\tilde{p},
\\
\partial_{t}\tilde{K}_{xz} + \partial_{x}\tilde{p}
&=
-\tilde{K}_{xz}
+
\frac{1}{\mathcal{R}}\nabla^{2}\tilde{K}_{xz}, \\ 
\partial_{t}\tilde{K}_{zz} + 2\,\partial_{z}\tilde{p}
&=
\tilde{\gamma}_{\varepsilon}\,\tilde{K}_{zz}
+
\frac{1}{\mathcal{R}}\nabla^{2}\tilde{K}_{zz}.
\end{align}
where
\begin{align}
\nabla^{2} = \partial_{z}^{2}+\partial_{x}^{2},
\qquad
\tilde{\gamma}_{\varepsilon} = \frac{1-\gamma-\varepsilon}{1+\varepsilon}.
\end{align}
Assuming normal modes of the form 
\begin{align}
    \tilde{p}
    & = 
    P \cos(nz) \cos(kx) e^{\sigma t} \\
    \tilde{K}_{xz}
    & = 
    U_{x} \cos(nz) \sin(kx) e^{\sigma t}\\ 
    \tilde{K}_{zz}
    & = 
    U_{z} \sin(nz) \cos(kx) e^{\sigma t}.
\end{align}
The system becomes 
\begin{align}
\mathcal{M} \cdot \mathcal{X} =  0, 
\end{align}
where $\mathcal{X} = [P, U_{x}, U_{z}]^{T}$ and 
\begin{align}
\mathcal{M} = \left[\begin{array}{ccc} 
\mathcal{L} & k & n \\ 
-k & \mathcal{D} & 0 \\ 
-2n & 0 & \mathcal{D} -(1+\tilde{\gamma}_{\varepsilon})
\end{array}\right].
\end{align}
with $\mathcal{L} = \kappa a^{2} + n^{2}$ and $\mathcal{D} = \sigma + \tfrac{k^{2} + n^{2}}{\mathcal{R}} + 1 $. The polarization (eigen)vector satisfies 
\begin{align}
    \mathcal{M}\cdot \left[\begin{array}{c} 
-n\mathcal{D}  \\ 
-n k  \\ 
k^{2} + \mathcal{L}\mathcal{D} 
\end{array}\right] = \left[\begin{array}{c} 
0  \\ 
0  \\ 
\det(\mathcal{M}) 
\end{array}\right].
\end{align}
The dispersion relation follows from the determinant condition, $\det(\mathcal{M}) = 0$, which has purely real solutions for $\sigma$.

\section{Numerical methods}

The system is discretized in Firedrake \cite{FiredrakeUserManual} on a periodic rectangular mesh of $N_x \times N_z$ quadrilateral cells, periodic in $x$ and bounded by Dirichlet walls in $z$. Pressure is represented in continuous $Q_1$; each independent component of $K$ lives in a discontinuous $Q_1$ (DG1) space, with the diffusion term $\eta\,\nabla^2 K$ treated by a symmetric interior penalty (SIPG) form with penalty parameter $\sigma_\mathrm{IP} = 6$ and a Nitsche-type imposition of the Dirichlet conditions at the $z$-walls. Pressure and conductivity are advanced monolithically by backward Euler with geometric step adaptation ($\Delta t \leftarrow 1.05\,\Delta t$ on success, $0.5\,\Delta t$ on Newton failure; $\Delta t \in [10^{-4}, 5\times 10^{-2}]$), solved via PETSc SNES (Newton with $\ell_2$ line search; $\mathrm{rtol} = 10^{-8}$, $\mathrm{atol} = 10^{-12}$) and an LU Jacobian solve through MUMPS \cite{Amestoy2000}. Because backward Euler does not preserve positive semi-definiteness of $K$, we apply a closed-form pointwise eigenvalue projection onto the PSD cone with floor $10^{-10}$, both as an SNES update callback at every Newton iterate and once more after each accepted step.

\section{Energy decay}

The energy of the gradient-flow PDE system is
\begin{align}
    E_{\text{total}} = E_{\text{transport}} + E_{\text{metabolic}} + E_{\text{diffusion}}
\end{align}
with
\begin{align}
    E_{\text{transport}}  &= \int_{M} \nabla p \cdot (K+ \kappa I) \cdot \nabla p \,\mathrm{d}x\,\mathrm{d}z \\
    E_{\text{metabolic}}  &= \frac{1+\varepsilon}{\gamma} \int_{M} f(\|K\|)^{\frac{\gamma}{2-\gamma}} \,\mathrm{d}x \,\mathrm{d}z \\
    E_{\text{diffusion}}  &= \frac{1}{2 \mathcal{R}} \int_{M} \|\nabla K\|^2 \,\mathrm{d}x \,\mathrm{d}z.
\end{align}
By construction, $E_{\text{total}}$ is monotonically non-increasing along
trajectories.

\begin{figure*}
    \centering
    \includegraphics[width=\textwidth]{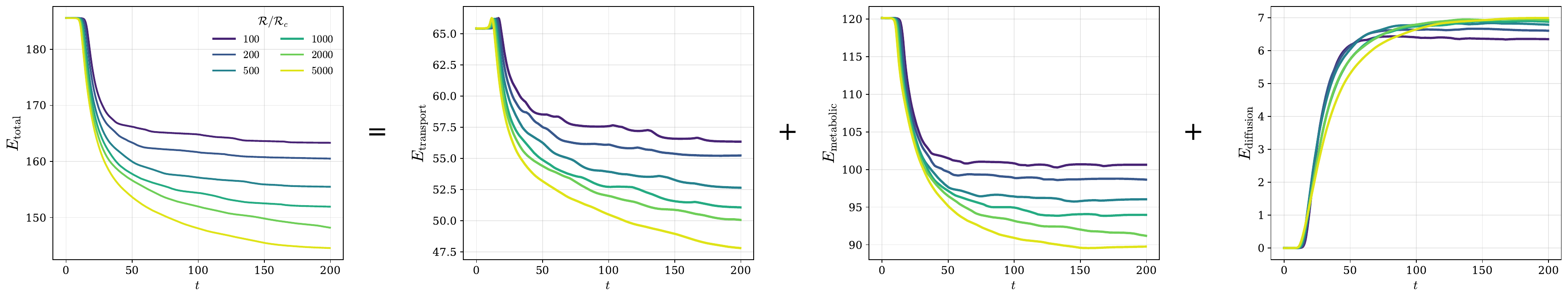}
    \caption{Energy decomposition of the continuum network-adaptation
    dynamics. Time evolution of the total energy $E_\mathrm{total}$ and
    its three constituents---transport $E_\mathrm{transport}$, metabolic
    $E_\mathrm{metabolic}$, and diffusion $E_\mathrm{diffusion}$---for
    the six supercriticalities reported in the main text.}
    \label{fig:energies}
\end{figure*}

Figure~\ref{fig:energies} shows the time evolution of each component
for the six supercriticalities reported in the main text. During the
linear exponential-growth phase ($t \lesssim t_\text{peak}$), all
components remain essentially flat: unstable perturbations are small
compared to the background. At saturation, $E_\text{transport}$ spikes
upward as the pressure field reorganizes around the freshly nucleated
channels and then drops as the channels mature; $E_\text{metabolic}$
decreases as conductivity is concentrated into channel cores;
$E_\text{diffusion}$, near zero in the homogeneous state, rises sharply
because $|\nabla K|^2$ is finite at channel edges. In the coarsening
phase, $E_\text{transport}$ and $E_\text{metabolic}$ exhibit jagged
step-like exchanges as individual channels merge or terminate, each
event redistributing material between the two contributions while
preserving the monotone decrease of $E_\text{total}$.

The $E_\text{diffusion}$ plateau visible at late times is the diffusive
anomaly discussed in the main text. Its dimensional form follows from
the integral: a channel of peak conductivity $K_\text{ch}$, width $w$,
and spacing $\lambda$ contributes $\sim K_\text{ch}^2 L_z / w$ to
$\int |\nabla K|^2 \,\mathrm{d}V$, and with $N \sim L_x/\lambda$ such
channels, $E_\text{diffusion} \sim L_x L_z K_\text{ch}^2/(\mathcal{R}\,
w\, \lambda)$. The plateau therefore identifies
$\mathcal{E} = K_\text{ch}^2/(\mathcal{R} w \lambda)$ as the relevant
scaling invariant.

\end{document}